\documentclass[journal]{IEEEtran}
\IEEEoverridecommandlockouts

\ifCLASSINFOpdf
\else
\fi

\usepackage{cite}
\usepackage{amsmath,amssymb,amsfonts}
\usepackage[linesnumbered,algoruled,boxed,lined]{algorithm2e}

\usepackage{graphicx}
\usepackage{textcomp}
\usepackage{xcolor}
\usepackage{multirow}
\usepackage{bm}
\usepackage{booktabs}
\usepackage{tabularx}
\usepackage[caption=false]{subfig}
\DeclareSubrefFormat{parens}{#1(#2)}
\usepackage{soul}

\allowdisplaybreaks[4]

\begin{document}

\pagestyle{empty} 

\title{LLM-guided DRL for Multi-tier LEO Satellite Networks with Hybrid FSO/RF Links}

\author{Jiahui Li, 
        Geng Sun,~\IEEEmembership{Senior Member,~IEEE,}
        Zemin Sun, 
        Jiacheng Wang, 
        Yinqiu Liu, 
        Ruichen Zhang, \\
        Dusit Niyato,~\IEEEmembership{Fellow,~IEEE,}
        Shiwen Mao,~\IEEEmembership{Fellow,~IEEE}
        

    \thanks{
    \par Jiahui Li and Zemin Sun are with the College of Computer Science and Technology, Jilin University, Changchun 130012, China (E-mails: lijiahui@jlu.edu.cn; sunzemin@jlu.edu.cn). 
    
    \par Geng Sun is with the College of Computer Science and Technology, Jilin University, Changchun 130012, China, and also with the Key Laboratory of Symbolic Computation and Knowledge Engineering of Ministry of Education, Jilin University, Changchun 130012, China. He is also with the College of Computing and Data Science, Nanyang Technological University, Singapore 639798 (E-mail: sungeng@jlu.edu.cn). 
    
    \par Jiacheng Wang, Yinqiu Liu, Ruichen Zhang, and Dusit Niyato are with the College of Computing and Data Science, Nanyang Technological University, Singapore 639798 (E-mails: jiacheng.wang@ntu.edu.sg; yinqiu001@e.ntu.edu.sg; ruichen.zhang@ntu.edu.sg; dniyato@ntu.edu.sg). 

    \par Shiwen Mao is with the Department of Electrical and Computer Engineering, Auburn University, Auburn, AL 36849-5201 USA (e-mail: smao@ieee.org).

    \par \textit{(Corresponding author: Geng Sun.)}
    }

}

\IEEEtitleabstractindextext{%
\begin{abstract}
Despite significant advancements in terrestrial networks, inherent limitations persist in providing reliable coverage to remote areas and maintaining resilience during natural disasters.  Multi-tier networks with low Earth orbit (LEO) satellites and high-altitude platforms (HAPs) offer promising solutions, but face challenges from high mobility and dynamic channel conditions that cause unstable connections and frequent handovers. In this paper, we design a three-tier network architecture that integrates LEO satellites, HAPs, and ground terminals with hybrid free-space optical (FSO) and radio frequency (RF) links to maximize coverage while maintaining connectivity reliability. This hybrid approach leverages the high bandwidth of FSO for satellite-to-HAP links and the weather resilience of  RF for HAP-to-ground links. We formulate a joint optimization problem to simultaneously balance downlink transmission rate and handover frequency by optimizing network configuration and satellite handover decisions. The problem is highly dynamic and non-convex with time-coupled constraints. To address these challenges, we propose a novel large language model (LLM)-guided truncated quantile critics algorithm with dynamic action masking (LTQC-DAM) that utilizes dynamic action masking to eliminate unnecessary exploration and employs LLMs to adaptively tune hyperparameters. Simulation results demonstrate that the proposed LTQC-DAM algorithm outperforms baseline algorithms in terms of convergence, downlink transmission rate, and handover frequency. We also reveal that compared to other state-of-the-art LLMs, DeepSeek delivers the best performance through gradual, contextually-aware parameter adjustments.
\end{abstract}

}

\maketitle

\IEEEdisplaynontitleabstractindextext
\IEEEpeerreviewmaketitle

%
\section{Introduction}
\label{sec:introduction}

\par With the rapid advancement of wireless communication technologies, terrestrial networks, including sixth-generation (6G) networks, have made significant progress in providing high-quality communication services. However, current terrestrial network architectures still face inherent challenges in delivering coverage to remote and underserved areas and maintaining resilience during natural disasters~\cite{Wang2024}. To overcome these limitations, non-terrestrial networks based on satellites have evolved beyond their traditional roles in global positioning, weather monitoring, and remote sensing to become integral components of modern communication infrastructures, substantially enhancing network coverage and disaster response capabilities~\cite{Wang2025}.

\par Low Earth orbit (LEO) satellite constellations, comprising thousands of satellites, play a particularly crucial role in this evolution due to their lower transmission delays compared to medium Earth orbit and geostationary Earth orbit satellites~\cite{Choi2024}. Despite their advantages, LEO satellites face significant challenges in establishing reliable direct communications with ground terminals due to their high mobility, relatively limited power resources, and the considerable path loss over long distances. These limitations often result in unstable connections, frequent handovers, and reduced quality of service when LEO satellites attempt to communicate directly with ground users~\cite{Toka2024}.

\par To address these challenges, high-altitude platforms (HAPs), operating at stratospheric altitudes of approximately 20 km, present a compelling solution to bridge this gap between LEO satellites and ground users~\cite{Elmahallawy2024}. By functioning as intermediate relay nodes, HAPs can significantly enhance the communication quality by providing a more stable link due to their quasi-stationary nature compared to rapidly moving LEO satellites. Furthermore, HAPs enjoy superior line-of-sight conditions to both satellites above and ground users below, effectively creating a multi-tier architecture that can overcome many of the aforementioned limitations inherent in direct LEO-to-ground communications~\cite{Sun2024}.

\par In such integrated satellite-HAP-terrestrial networks, the strategic integration of communication technologies for different segments is crucial for system performance. Specifically, free-space optical (FSO) communication offers an ideal solution for satellite-to-HAP links due to its high bandwidth capacity, license-free spectrum, and excellent performance in the absence of atmospheric disturbances at higher altitudes~\cite{Samy2023}. Meanwhile, radio frequency (RF) technology remains better suited for HAP-to-ground links as it provides greater resilience against adverse weather conditions such as cloud coverage, fog, and precipitation that would severely attenuate optical signals~\cite{Sun2024}. This hybrid FSO/RF architecture leverages the strengths of each technology while mitigating their respective weaknesses, creating a more robust and efficient communication system.

\par However, designing and optimizing such a multi-tier hybrid satellite downlink communication system presents several significant challenges. \textit{Firstly}, the dynamic nature of LEO satellite movement creates an ever-changing network topology, which necessitates a sophisticated handover mechanism to maintain continuous connectivity~\cite{Wu2024a}. \textit{Secondly}, the heterogeneous nature of FSO and RF links introduces complex interdependencies in resource allocation decisions~\cite{Zeng2025}. \textit{Finally}, the quasi-stationary yet mobile nature of HAPs adds another layer of complexity to the system, requiring adaptive methods that can account for gradual position shifts and their impact on link quality and coverage footprint. These interrelated challenges call for an innovative approach that can effectively coordinate across multiple network tiers and communication technologies.

\par Accordingly, we aim to propose a novel online jointly handover and network configuration approach for the considered multi-tier hybrid satellite downlink communication system that directly addresses these challenges. The main contributions of this paper are summarized as follows:

\begin{itemize}
    \item \textit{Multi-tier Hybrid FSO/RF Satellite Downlink Communication Architecture:} We design a three-tier network architecture that integrates LEO satellites, HAPs, and ground terminals with hybrid FSO/RF links to maximize coverage while maintaining connectivity reliability. This architecture is particularly effective in remote areas, disaster scenarios, and regions with challenging terrain where traditional infrastructure is limited or compromised. To the best of our knowledge, this is the first work to jointly optimize the handover mechanism and network configuration in such a multi-tier hybrid communication system while considering the distinctive characteristics of each network layer.

    \item \textit{Dynamic Joint Optimization Problem with Cross-tier Dependencies:} We model the system to capture the complex interplay among LEO satellite mobility, HAP positioning, and channel dynamics across the multi-tier network. We find that dependencies exist between handover and resource allocation between LEO satellite and HAP tiers. As such, we formulate a joint optimization problem to simultaneously balance the downlink transmission rate and handover times by optimizing the network configuration and satellite handover decisions. The problem is highly dynamic and non-convex with time-coupled constraints, which requires a method with enhanced adaptability to manage temporal variations. 

    \item \textit{Large Language Model (LLM)-guided DRL-based Algorithm:} Traditional optimization approaches struggle with the dynamic and non-convex nature of the optimization problem. To address these challenges, we propose a novel deep reinforcement learning (DRL)-based algorithm, namely, LLM-guided truncated quantile critics algorithm with dynamic action masking (LTQC-DAM), to solve the optimization problem. Specifically, LTQC-DAM incorporates the dynamic action masking mechanism to eliminate unnecessary exploration and employs the LLM to adaptively tune the hyperparameters, thereby achieving fast and adaptive convergence. 
    
    \item \textit{Comprehensive Performance Analysis with Empirical Validation:} Simulation results demonstrate that the LTQC-DAM algorithm achieves faster convergence compared to various baseline algorithms. Moreover, LTQC-DAM improves the downlink transmission rates and achieves a 17.69\% improvement in satellite handover frequency. We also evaluate the impact of different LLM models (including DeepSeek, Qwen, Claude, ChatGPT, and Grok) on the LTQC-DAM algorithm, and reveal that DeepSeek delivers the best performance through gradual and contextually-aware parameter adjustments.

\end{itemize}

\par The rest of this paper is arranged as follows. Section \ref{sec:related_work} reviews the related works. Section \ref{sec:models_and_preliminaries} presents the models and problem. Section \ref{sec:algorithm} proposes the algorithm. Section \ref{sec:simulation_results_and_analysis} shows the simulation results, and Section \ref{sec:conclusion} concludes the paper.

%
%
\section{Related Work} 
\label{sec:related_work}

\par In this section, we briefly introduce related works to highlight our innovations and contributions in multi-tier hybrid FSO/RF satellite communication systems.

%
%
\subsection{FSO-based Satellite Network Architectures} 

\par Due to its exceptional bandwidth capabilities and minimal spectrum regulation constraints, FSO communication has emerged as a valuable technique for next-generation satellite communications~\cite{Chen2025, Elamassie2025}. For example, the authors in~\cite{Nguyen2024} discussed the design of blind key reconciliation schemes for LEO satellite systems based on FSO, which enables operation without a prior quantum bit-error rate estimation. Moreover, the authors in~\cite{Le2023} proposed a cross-layer design framework of error-control protocols with rate adaptation for FSO burst transmissions in HAP-aided space-air-ground integrated vehicular networks. In addition, the authors in~\cite{Le2022} proposed a hybrid automatic repeat request-based sliding window mechanism for the high data-rate and long-distance of FSO-based satellite systems. However, FSO communications encounter performance degradation due to atmospheric disturbances, primarily precipitation and fog, causing beam distortion and reduced optical power~\cite{Wu2024}.

\par To overcome these environmental limitations, hybrid FSO/RF transmission architectures have been extensively investigated to offer a more robust solution for practical deployment scenarios~\cite{Sun2024}. For instance, the authors in~\cite{Samy2023a} deployed a UAV as a relay in hybrid FSO/RF transmissions to mitigate the adverse effect of atmospheric turbulence and weather dependence. Moreover, the authors in~\cite{Shah2021} considered a hybrid FSO/RF communication between the ground station and the satellite, where the RF link improves the reliability of the FSO communications, and HAP improves the end-to-end transmission. In~\cite{Huang2021}, the authors deployed a HAP as a relay to assist uplink transmission of the space-air-ground network with RF and FSO links. However, these works predominantly concentrate on static communication link establishment while overlooking several dynamic aspects, such as satellite mobility patterns, inter-satellite handover complexities, and the non-autonomous drift of HAPs due to stratospheric winds.

%
%
\subsection{Optimization Metrics in Hybrid FSO/RF Satellite Systems} 

\par Optimizing performance metrics while maintaining service continuity represents a primary challenge in hybrid FSO/RF satellite systems. As such, the authors in~\cite{Ma2022} investigated the secrecy outage performance of the uplink transmission in the hybrid FSO/RF cooperative space-air-ground network and analyzed the performance using a stochastic geometry framework. Moreover, the authors in~\cite{Li2023a} investigated the performance of the hybrid FSO/RF reconfigurable intelligent surfaces (RIS)-assisted UAV relay systems and evaluated the outage probability, average bit-error rate, and ergodic capacity. In addition, the authors in~\cite{Bankey2023} proposed to use hybrid FSO/RF transmission to enhance the physical layer security of the downlink space-air-ground integrated network. However, these studies predominantly focus on static link reliability metrics without addressing the complexities of actual data transmission in dynamic satellite environments.

\par Furthermore, recent studies have elevated research focus toward optimizing transmission rates in hybrid FSO/RF space-air-ground communications. For example, the authors in~\cite{Nguyen2023} explored a novel solution for the hybrid FSO/RF HAP-based space-air-ground network to optimize the average transmission rate. In~\cite{Samy2024}, the authors proposed a hybrid space-air-ground FSO/RF transmission system with multiple HAP relays to optimize throughput, spectral efficiency, and bit-error rate. Moreover, the authors in~\cite{Lee2023} investigated the problem of forwarding packets between ground terminals using RF and FSO links to maximize the communication efficiency. Despite these advancements, a critical gap persists in addressing satellite handover procedures, which significantly transform transmission channel characteristics and ultimately determine real-world system viability in dynamic LEO constellations.

\subsection{Optimization Methods in FSO/RF Satellite Network} 

\par A variety of innovative optimization methods have been proposed to enhance FSO/RF LEO satellite network performance under challenging conditions. For example, in~\cite{Samy2023b}, the authors proposed the parallel FSO and RF transmissions to explore their complementary properties in beamwidth and bandwidth, and derived a sum capacity outage probability. Moreover, the authors in~\cite{Samy2024a} explored a hybrid single-hop FSO/RF transmission system and performed a capacity analysis in the presence of correlated turbulence. Despite these advancements, these approaches still exhibit limitations in dynamic adjustment capabilities and lack timely feedback mechanisms.

\par To address these limitations, researchers have begun incorporating learning-based optimization methods into hybrid FSO/RF satellite systems. The authors in~\cite{Wu2024} introduced a HAP-based integrated relay network with a hybrid FSO/RF transmission mode, incorporating UAVs equipped with RISs to dynamically reconfigure the propagation environment, and proposed a DRL-based framework to maximize the system ergodic rate. Moreover, the authors in~\cite{Ibrahim2024} implemented a supervised learning model to anticipate FSO attenuation to maximally maintain a high data rate in the RF/FSO links. Nevertheless, these methods often overlook the solution action space created by satellite constellations, fail to eliminate irrelevant actions, and disregard the random fluctuations in optimization objectives caused by channel variations. 

\par In summary, different from the existing works, we consider a multi-tier hybrid FSO/RF satellite downlink communication architecture. Accordingly, we aim to devise an online method with enhanced adaptability to manage random fluctuations caused by channel variations.

\section{Models and Problem Formulation}
\label{sec:models_and_preliminaries}

\par In this section, we first present the system overview. Then, we introduce the detailed models, including the satellite orbit, HAP mobility, network architecture, and handover models, to characterize the concerned objectives and key decision variables. The main notations are presented in Table \ref{tab:notations}.

\begin{table}[]
\centering
\caption{Main notations}
\label{tab:notations}
\begin{tabularx}{3.5in}{p{1.5cm}p{6.5cm}}
\toprule
\textbf{Symbol} & \textbf{Definition} \\ 
\midrule
$a[t]$ & Action at time slot $t$ \\
$\mathcal{C}_i$, $N_C$ & The $i$th ground user cluster and the total number of clusters \\
$\mathcal{H}$ & High-altitude platform (HAP) \\
$h_{SH}(t)$ & FSO channel gain between satellite and HAP at time $t$ \\
$h_{HC,i}(t)$ & Channel gain between HAP and the selected user of cluster $i$ at time $t$ \\
$\mathcal{L}$, $N_L$ & LEO satellite constellation set and the total number of satellites \\
$\mathcal{L}_t$ & Available LEO satellite set at time slot $t$ \\
$N_t$ & Number of satellite handovers from time slot 0 up to $t$ \\
$\boldsymbol{N}$ & Decision variables for subcarrier allocation \\
$\mathbf{n}(t)$ & Subcarrier allocation vector at time $t$ \\
$R_{FSO}(t)$ & Data transmission rate of the FSO link at time $t$ \\
$R_{RF_i}(t)$ & Data rate for the $i$th cluster at time $t$ \\
$R_{total}(t)$ & Effective total transmission rate at time $t$ \\
$r[t]$ & Reward at time slot $t$ \\
$s_t$ & Index of the connected LEO satellite at time slot $t$ \\
$\boldsymbol{S}$ & Decision variables for satellite selection \\
$\mathcal{S}, \mathcal{A}, \mathcal{P}, R, \gamma$ & MDP components: state space, action space, transition probabilities, reward function, discount factor \\
$s[t]$ & State at time slot $t$ \\
$\mathcal{T}$ & System timeline set $\{1,2,\ldots ,T\}$ \\
$\boldsymbol{U}$ & Decision variables for user selection within clusters \\
$\mathbf{u}(t)$ & User selection vector at time $t$ \\
$\mathbf{v}(t), \mathbf{p}(t)$ & Velocity and position vectors of the HAP at time $t$ \\
$\iota, \Omega, \omega, \varepsilon, \varrho, \nu$ & Orbital parameters: inclination angle, right ascension, argument of perigee, eccentricity, semi-major axis, true anomaly \\
$\eta, \zeta$ & Weighting factors for throughput maximization and handover minimization in the reward function \\
\bottomrule
\end{tabularx}
\end{table}

%
%
\subsection{System Overview}

%
\begin{figure}
  \centering
  \includegraphics[width=3.5in]{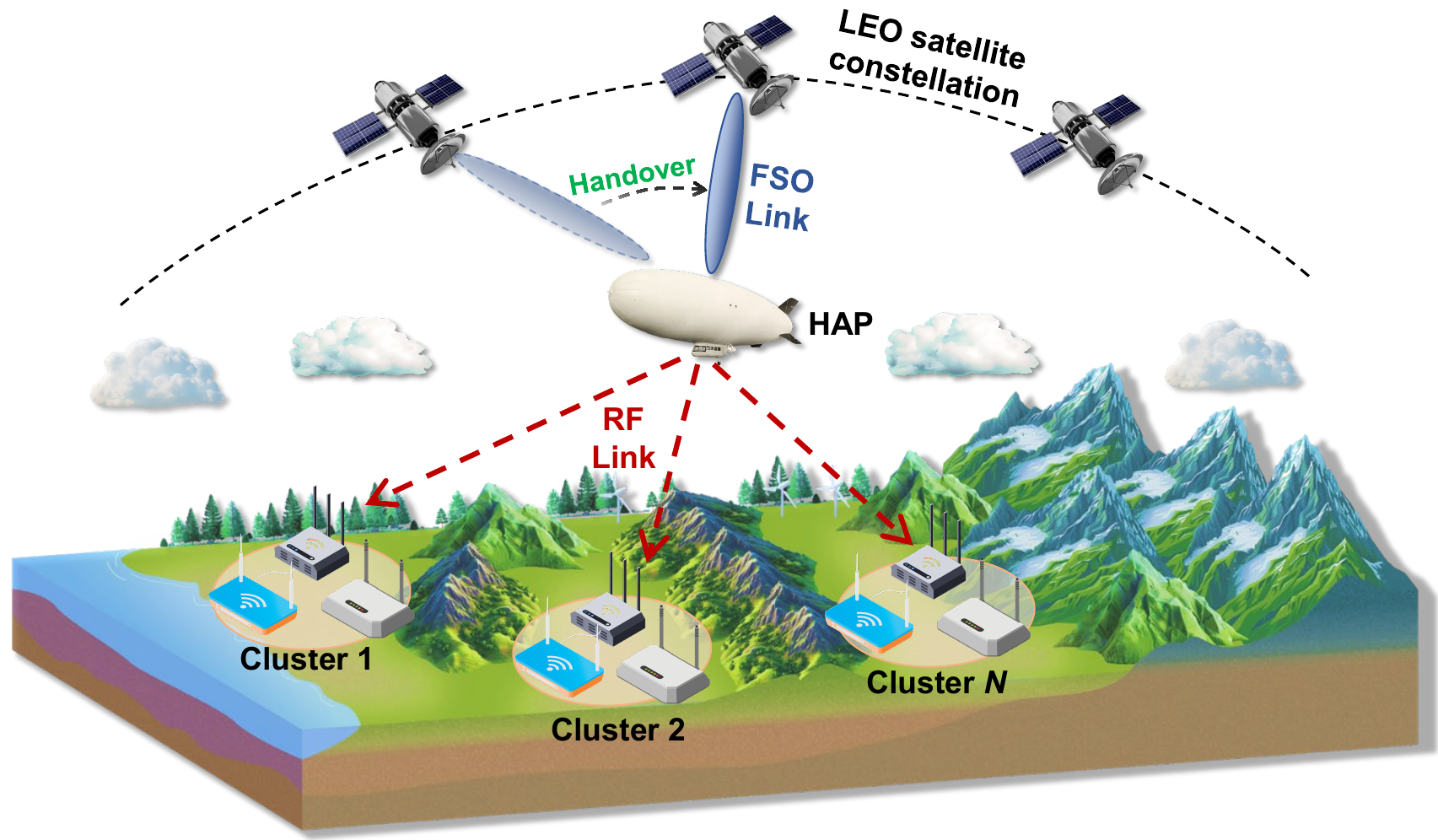}
  \caption{The considered multi-tier hybrid satellite downlink communication system composed of a LEO satellite constellation, an HAP, and multiple ground user clusters.}
  \label{fig:network-model}
\end{figure}

\par As shown in Fig.~\ref{fig:network-model}, we consider a multi-tier hybrid satellite downlink communication system composed of a LEO satellite constellation, an HAP, and multiple ground user clusters. Specifically, the LEO satellite constellation is represented as $\mathcal{L} = \{ 1, 2, \dots, N_L \}$, where $N_L$ is the total number of satellites. These satellites move in periodic orbits and are responsible for establishing communication links with the HAP, denoted as $\mathcal{H}$. Moreover, the HAP moves dynamically in the atmosphere and acts as an intermediary between the LEO satellites and the ground users. In addition, the ground users are divided into $N_C$ clusters due to the natural geographical environment, denoted as $\mathcal{C}_i$ for $i \in \{1, 2, \dots, N_C \}$, where the $i$th cluster $\mathcal{C}_i$ consists of multiple users $\{1, 2, \dots, N_{U,i}\}$ that can communicate internally. 

\par During each mission cycle, one available satellite from the constellation $\mathcal{L}$ will establish an FSO downlink with the HAP $\mathcal{H}$ to transmit data. Then, the HAP simultaneously communicates with multiple ground user clusters using OFDM technology for efficient resource allocation. These processes involve dynamic satellite handover and subcarrier allocation to ground clusters. As such, we consider a discrete-time system evolving over the timeline $\mathcal{T} = \{t|1,2,\ldots, T \}$. At each time slot, a subset of the LEO satellites has suitable angles to perform FSO downlink, and the available LEO satellite set at the $t$th time slot is denoted as $\mathcal{L}_t \subseteq \mathcal{L}$. The LEO constellation needs to select one LEO satellite to connect with the HAP, and we denote the index of the connected LEO satellite at the $t$th time instant as $s_t$.

\par Without loss of generality, we consider a 3D Cartesian coordinate system. The position of the $i$th LEO satellite $L_i$ is given by $(x^{L}_i(t), y^{L}_i(t), z^{L}_i(t))$ at time $t$. Similarly, the position of the HAP is represented as $(x^{H}(t), y^{H}(t), z^{H}(t))$, and the position of the $k$th user in the $i$th ground cluster is denoted as $(x^{C}_{i,k}, y^{C}_{i,k}, z^{C}_{i,k})$. These positions are updated periodically as the satellites move along their orbits and the HAP adjusts its position according to its mobility model. In the following subsections, we will describe the detailed models for the satellite motion, network architecture, HAP mobility, and handover mechanisms.

%
%
\subsection{LEO Satellite Orbit}

\par LEO satellites orbit Earth at altitudes ranging from 500 to 2000 kilometers, which allows them to complete an orbit in a short period~\cite{Pan2023}. The orbital parameters of a LEO satellite can be described by the tuple $<\iota, \Omega, \omega, \varepsilon, \varrho, \nu>$~\cite{Montenbruck2002}, \textit{i.e.,}

\begin{itemize}
  \item \textit{Inclination Angle ($\iota$):} The angle between the orbital plane and the equator. 
  \item \textit{Right Ascension of Ascending Node ($\Omega$):} The angle between the vernal equinox and the point where the satellite crosses the equatorial plane.
  \item \textit{Argument of the Perigee ($\omega$):} The angle between the ascending node and the perigee, the point closest to Earth.
  \item \textit{Eccentricity ($\varepsilon$):} Describes the shape of the orbit. 
  \item \textit{Semi-Major Axis ($\varrho$):} Half of the longest diameter of the elliptical orbit.
  \item \textit{True Anomaly ($\nu$):} The angle between the perigee direction and the current position of the satellite.
\end{itemize}

\par To simplify the analysis, we consider the circular orbits for the LEO satellites~\cite{Deng2021}. Thus, $\varepsilon = 0$, and the semi-major axis $\varrho$ is the same as the orbital radius $H_i = h_i + R_e$, where $h_i$ is the altitude of the satellite, and $R_e$ is the radius of Earth. The angular velocity $\varpi_i$ of the satellite is given by $\varpi_i= \sqrt{GM_e/H_i^3}$, where $G$ is the gravitational constant and $M_e$ is the mass of Earth, which allows the orbital period $\tau_i$ to be expressed as $\tau_i = 2 \pi / \varpi_i$. 

\par As such, the 3D position of satellite $i \in \mathcal{L}$ at time slot $t$ is denoted as $(x^{L}_i(t), y^{L}_i(t), z^{L}_i(t))$ and is determined by
\begin{equation} \label{eq:orbit}
\begin{aligned}
&x^{L}_{i}(t) = H_i \left(\cos (\omega_{i}^{t}+\nu_i) \cos \Omega_i-\sin (\omega_{i}^{t}+\nu_i) \cos \iota_i \sin \Omega_i \right), \\
&y^{L}_{i}(t) = H_i \left(\cos (\omega_{i}^{t}+\nu_i) \sin \Omega_i+\sin (\omega_{i}^{t}+\nu_i) \cos \iota_i \cos \Omega_i \right), \\
&z^{L}_{i}(t) = H_i \left( \sin (\omega_{i}^{t}+\nu_i) \sin \iota_i \right),
\end{aligned}
\end{equation}

\noindent where $\omega_{i}(t) = \omega_i^{init} + (t \varpi_i \mod \tau_i)$ represents the time-varying argument of perigee for satellite $i$. Following this, at each time slot $t$, the system identifies the subset of satellites $\mathcal{L}_t \subseteq \mathcal{L}$ that have favorable communication conditions with the HAP, based on visibility, elevation angle, and potential signal quality. The determination of $\mathcal{L}_t$ depends on the orbital parameters and positions of both the satellites and the HAP. This available satellite set $\mathcal{L}_t$ forms the basis for satellite selection decisions. Simultaneously, the position of each satellite continuously changes over its orbital period $\tau_i$, creating a dynamic communication environment. Such dynamic behavior necessitates adaptive decision-making for satellite handover optimization.

%
%
\subsection{HAP Mobility Model}
\label{ssec: HAP}

\par The HAP operates at an altitude of approximately 20 km and moves dynamically in the atmosphere~\cite{Mashiko2025}. As such, we propose to employ the Gauss-Markov mobility model to characterize the mobility pattern of the HAP~\cite{Sun2025}, thereby capturing the temporal dependency in the HAP movement. Accordingly, the three-dimensional velocity vector $\mathbf{v}(t) = [v_x(t), v_y(t), v_z(t)]$ and position vector $\mathbf{p}(t) = [x^{H}(t), y^{H}(t), z^{H}(t)]$ of the HAP are updated as follows:
\begin{equation}
\begin{aligned}
\mathbf{v}(t+1) = \alpha \cdot \mathbf{v}(t) + (1-\alpha) \cdot \mathbf{\mu}_v + \sqrt{1-\alpha^2} \cdot \mathbf{\sigma} \cdot \mathbf{n}(t),
\end{aligned}
\end{equation}

\begin{equation}
\begin{aligned}
\mathbf{p}(t+1) = \mathbf{p}(t) + \mathbf{v}(t+1) \cdot \Delta t,
\end{aligned}
\end{equation}

\noindent where $\alpha$ is the memory parameter that determines how much the velocity depends on its previous value, $\mathbf{\mu}_v = [\mu_{v_x}, \mu_{v_y}, \mu_{v_z}]$ is the mean velocity vector, $\mathbf{\sigma} = [\sigma_{v_x}, \sigma_{v_y}, \sigma_{v_z}]$ represents the standard deviation of velocity in each dimension, and $\mathbf{n}(t) = [n_x(t), n_y(t), n_z(t)]$ is a vector of independent Gaussian random variables with zero mean and unit variance. Note that we consider that the movement of HAP is bounded within the specific operational area. When the HAP reaches these boundaries, it implements a reflection strategy by reversing the corresponding velocity component.

%
%
\subsection{Network Models}

\par In this section, we present the network models, including satellite-to-HAP and HAP-to-ground cluster communication tiers.

\subsubsection{FSO-based Satellite-to-HAP Communication Model}

\par In the designed multi-tier hybrid satellite downlink communication system, the communication process between the LEO satellite constellation $\mathcal{L}$ and the HAP $\mathcal{H}$ utilizes FSO communication due to its direct line-of-sight path and low interference in atmospheric conditions. As such, the FSO channel gain between a satellite and the HAP at time $t$ can be expressed as follows~\cite{Wu2024}:
\begin{equation}
h_{SH} (t) = h_l \cdot h_a(t),
\end{equation}

\noindent where $h_l$ represents the link loss component which accounts for transmit antenna gain ($G_T$), receive antenna gain ($G_R$), free-space loss ($A_{FS}$), atmospheric attenuation ($A_{ATM}$), lenses loss ($L_{loss}$), and system margin ($M_S$), which is given by 
\begin{equation}
h_l = \frac{1}{2}(G_T + G_R - A_{FS} - A_{ATM} - L_{loss} - M_S).
\end{equation}

\par Moreover, the channel fading component $h_a(t)$ follows a Gamma-Gamma probability distribution $\Gamma(\alpha, \beta)$ with shape parameters $\alpha$ and $\beta$, which is a widely adopted statistical model specifically designed to capture the impacts of fading caused by atmospheric turbulence on optical wave propagation. 

\par Following this, the signal-to-noise ratio (SNR) at the HAP can be expressed as follows~\cite{Wu2024}:
\begin{equation}
\gamma_H(t) = \frac{P_{FSO} \eta_{OE}^2 h_{EGC}^2(t)}{N_A N_q} = \bar{\gamma}_H h_{EGC}^2(t),
\end{equation}
\noindent where $P_{FSO}$ is the satellite transmit power, $\eta_{OE}$ is the optical-to-electrical conversion efficiency, $h_{EGC}(t) = \sum_{q=1}^{N_A} h_{SH}(t)$ represents the scalar channel fading of the receive aperture ensemble, $N_A$ is the number of receiver apertures to collect optical signals, $N_q$ is the noise power, and $\bar{\gamma}_H = \frac{P_{FSO}\eta_{OE}^2}{N_A N_q}$ represents the average SNR.

\par As such, the corresponding data transmission rate between the satellite and the HAP is given by
\begin{equation}
R_{FSO}(t) = B_{FSO} \log_2 \left( 1 + \gamma_H(t) \right),
\end{equation}

\noindent where $B_{FSO}$ denotes the bandwidth of the FSO-based link.

\subsubsection{OFDM-based HAP-to-Ground Cluster Communication Model}

\par The communications between the HAP and the ground user clusters $\mathcal{C}_i$ in the designed multi-tier hybrid satellite downlink communication system are based on radio frequency and involve orthogonal frequency-division multiplexing (OFDM) technology. Specifically, the HAP is equipped with the total of $N_S$ OFDM subcarriers that can be flexibly allocated among the $N_C$ ground clusters. For each cluster, the HAP selects one user as the communication endpoint based on channel conditions, typically the user with the best channel quality.

\par Following this, we employ the Nakagami-$m$ fading to model the channel between the HAP and ground users, according to~\cite{Wu2024}. Thus, this channel gain between the HAP and the selected user of the cluster $\mathcal{C}_i$ is given by~\cite{Li2024}
\begin{equation}
h_{HC,i}(t) = C_{HC,i}(t) g_{HC,i}(t),
\end{equation}

\noindent where $g_{HC,i}(t)$ denotes the Nakagami-$m$ distributed random variable with channel fading severity parameter $m$, and $C_{HC,i}(t)$ represents the loss component of the RF signal from the HAP to the user, which can be expressed as
\begin{equation}
\begin{aligned}
    C_{HC, i}&(t)  = \\ & G_{HC} + R_i + \frac{1}{2}(20 \lg\lambda_F - 10\eta\lg d_{HC,i}(t) - 20 \lg 4\pi),
\end{aligned}
\end{equation}

\noindent where $G_{HC}$ and $R_i$ represent the HAP transmit antenna gain and the IoT device receive antenna gain, respectively, $\eta$ denotes the path loss coefficient, $\lambda_F$ is the FSO channel wavelength, and $d_{HC, i}(t)$ represents the signal transmission distance from the HAP to the selected user of the cluster $\mathcal{C}_i$ at time $t$.

\par Following this, the data rate for the $i$th cluster can be expressed as 
\begin{equation}
R_{RF_i}(t) = n_i(t) \cdot \frac{B_{RF}}{N_S} \log_2 \left( 1 + \frac{P_{RF} |h_{HC,i}(t)|^2}{(B_{RF}/N_S) \cdot \sigma_C^2} \right),
\end{equation}
\noindent where $n_i(t)$ is the number of subcarriers allocated to cluster $i$ at time $t$, $B_{RF}$ is the total RF bandwidth, $P_{RF}$ is the transmit power, and $\sigma_C^2$ is the noise power spectral density.

\par Moreover, due to the decode-and-forward (DF) protocol applied at the HAP, the total data rate across all clusters is constrained by the FSO link rate between the satellite and HAP, \textit{i.e.,}
\begin{equation}
\sum_{i=1}^{N_C} R_{RF_i}(t) \leq R_{FSO}(t).
\end{equation}

\noindent Note that this constraint ensures that the HAP cannot distribute more data to ground clusters than it receives from the satellite. As such, there is a significant interdependency between LEO satellite handover events and resource allocation in this multi-tier network architecture. When satellite handover occurs, the FSO channel characteristics ($h_{SH}(t)$) undergo substantial changes, directly affecting the achievable data rate $R_{FSO}(t)$ and consequently limiting the total resources available for distribution to ground clusters through the constraint $\sum_{i=1}^{N_C} R_{RF_i}(t) \leq R_{FSO}(t)$. This cascading effect requires the HAP to dynamically adjust the subcarrier allocation parameters $n_i(t)$ during handover transitions, which highlights the need for joint optimization of handover and resource allocation.

%
%
\subsection{Satellite Handover Model}

\par At each time slot $t \in \mathcal{T}$, the LEO satellite constellation $\mathcal{L}$ determines which satellite is used to establish a connection to the HAP throughout the FSO link. If the satellite constellation assigns a satellite, the satellite will track the HAP throughout the time slot. Moreover, at the start of each time slot, the LEO satellite constellation makes the decision to maintain the current connection or switch to a new one. As such, the sequence of satellite handover at each time slot can be represented by $\boldsymbol{S} = \{s_t | t \in \mathcal{T}, s_t \in \mathcal{L}_t \}$. Thus, this sequence affects both the end-to-end downlink transmission rate and the frequency of satellite handovers.

%
%
\subsection{Problem Formulation}

\par In this work, we aim to optimize the downlink communication in a multi-tier system involving the LEO satellite constellation, the HAP, and multiple ground user clusters. Our primary objective is to maximize the end-to-end downlink transmission rate from the satellite to the ground user clusters, while minimizing the frequency of satellite handovers to reduce overhead and ensure seamless communication.

\par To achieve these objectives, we introduce three parts of decision variables as follows. \textit{(i)} $\boldsymbol{N} = \{n_i(t) | i \in \{1,2,\ldots,N_C\}, t \in \mathcal{T}, n_i(t) \in \{0,1,\ldots,N_S\} \}$, representing the number of OFDM subcarriers allocated to each ground cluster $i$ at each time slot $t$. \textit{(ii)} $\boldsymbol{S} = \{s_t | t \in \mathcal{T}, s_t \in \mathcal{L}_t \}$, representing the index of the selected satellite at each time slot. \textit{(iii)} $\boldsymbol{U} = \{u_{i,t} | i \in \{1,2,\ldots,N_C\}, t \in \mathcal{T}, u_{i,t} \in \{1,2,\ldots,N_{U,i}\} \}$, representing the selected user from each ground user cluster for communication.

\par \textit{Optimization Objective 1:} The first optimization objective is to maximize the total end-to-end downlink transmission rate from the LEO satellite through the HAP to the ground users. For each time slot $t$, the FSO link between the selected satellite and HAP achieves a rate of $R_{FSO}(t)$, while the RF links to the various clusters achieve rates of $R_{RF_i}(t)$ based on the selected subcarrier allocation and user selection. This optimization objective can be expressed as
\begin{equation}
    f_1(\boldsymbol{N}, \boldsymbol{S}, \boldsymbol{U}) = \sum_{t \in \mathcal{T}} R_{total}(t),
\end{equation}
where $R_{total}(t)$ represents the effective total transmission rate at time $t$. Due to the flow conservation constraint, this rate is determined by
\begin{equation}
    R_{\text{total}}(t) = \min \left\{ R_{FSO}(t), \sum_{i=1}^{N_C} R_{RF_i}(t) \right\}.
\end{equation}

\par Note that if the FSO link can support higher rates than the RF links ($R_{FSO}(t) \geq \sum_{i=1}^{N_C} R_{RF_i}(t)$), then each cluster receives its full rate $R_{RF_i}(t)$. However, if the FSO link becomes the bottleneck ($R_{FSO}(t) < \sum_{i=1}^{N_C} R_{RF_i}(t)$), the available FSO capacity will be proportionally distributed among the clusters, \textit{i.e.},
\begin{equation}
    R'_{RF_i}(t) = \frac{R_{RF_i}(t)}{\sum_{j=1}^{N_C} R_{RF_j}(t)} \cdot R_{FSO}(t).
\end{equation}

\par \textit{Optimization Objective 2:} The second optimization objective is to minimize the number of satellite handovers. This is because frequent satellite handovers can introduce communication delays and additional overhead. Mathematically, the number of satellite handovers at time slot $t$, denoted as $N_t$, is updated as follows:
\begin{equation}
    N_{t+1} = \begin{cases}
N_t, & \text{if} \quad  s_t = s_{t+1}, \\
N_t + 1, & \text{if} \quad  s_t \neq s_{t+1}.
\end{cases}
\end{equation}
Thus, the second optimization objective is to minimize the total number of satellite handovers, \textit{i.e.}, 
\begin{equation}
    f_2(\boldsymbol{S}) = N_{\mathcal{T}}.
\end{equation}

\par To jointly optimize the downlink transmission rate and minimize the satellite handovers, we formulate the optimization problem as follows:
\begin{subequations}
  \label{eq:formulation}
  \begin{align}
    (\mathrm{P1}): {\underset{\boldsymbol{N}, \boldsymbol{S}, \boldsymbol{U}}{\text{max}}} &  F=(f_1, -f_2),\\
    \text{s.t.} \quad 
    & n_i(t) \in \{0,1,\ldots,N_S\}, \\
    & \quad \quad \left(\forall i \in \{1,2,\ldots,N_C\}, \forall t \in \mathcal{T}\right), \label{eq:const1} \\
    & s_t \in \mathcal{L}_t, \quad \forall t \in \mathcal{T}, \label{eq:const2} \\
    & u_{i,t} \in \{1,2,\ldots,N_{U,i}\}, \\ & \quad \quad \left(\forall i \in \{1,2,\ldots,N_C\}, \forall t \in \mathcal{T}\right), \label{eq:const3} \\
    & \sum_{i=1}^{N_C} n_i(t) = N_S, \quad \forall t \in \mathcal{T}. \label{eq:const4}
  \end{align}
\end{subequations}

\noindent Note that constraint~\eqref{eq:const1} ensures that the number of subcarriers allocated to each cluster is within the valid range. Constraint~\eqref{eq:const2} ensures that the selected satellite is from the set of visible satellites at each time slot. Constraint~\eqref{eq:const3} governs the user selection within each cluster, where the selected user must be among the $N_{U,i}$ users in cluster $i$. Finally, constraint~\eqref{eq:const4} ensures that the total number of subcarriers allocated across all clusters equals the available number of subcarriers $N_S$.

\par The optimization problem (P1) is challenging to solve analytically due to the time-varying channel conditions and the complex mobility patterns of both satellites and HAP. Such high dynamicity necessitates an adaptive and responsive algorithm capable of real-time decision-making. Moreover, the sequential decision-making process in the satellite handover procedure creates strong temporal dependencies between decisions made at different time slots. Capturing such time dependencies and developing a tractable decision framework remains a significant technical challenge. Therefore, we seek to design a novel DRL-based method that is well-suited for sequential decision-making in complex and dynamic environments to address this problem.

%
\section{LLM-guided DRL-based approach} 
\label{sec:algorithm}

\par In this section, we propose a DRL-based method to solve the formulated optimization problem. To this end, we first show the motivations for using DRL and reformulate the problem as a Markov decision process (MDP). Then, we introduce the proposed algorithm with key improvements.

\subsection{Preliminaries of DRL and MDP Formulation}
\label{ssec:DRL}

\par As aforementioned, the formulated problem is inherently dynamic and stochastic, involving continuously changing system parameters such as the positions of LEO satellites, user channel conditions, and the HAP mobility. This results in a highly uncertain environment where decisions regarding satellite handover, subcarrier allocation, and user selection must be made sequentially in real-time. Traditional optimization techniques, such as convex optimization and evolutionary algorithms, struggle to adapt to such dynamic environments, particularly when system conditions fluctuate rapidly.

\par In such cases, DRL offers a powerful solution due to its fundamental ability to learn optimal policies through direct interaction with complex environments~\cite{Li2024a}. Different from the traditional methods requiring complete mathematical characterization of the system, DRL excels in environments with high-dimensional state spaces, nonlinear dynamics, and uncertain transitions. Moreover, by leveraging deep neural networks as function approximators, DRL can efficiently handle the complexities introduced by time-varying states and sequential decision processes~\cite{Zhang2025}. Thus, based on the DRL frameworks, we can develop a robust algorithm that dynamically adjusts decision variables to maximize overall system performance.

\par To solve this problem via DRL-based frameworks, we first reformulate the optimization problem (P1) as a Markov Decision Process (MDP). An MDP is defined by the tuple $(\mathcal{S}, \mathcal{A}, \mathcal{P}, R, \gamma)$, where $\mathcal{S}$ is the state space, $\mathcal{A}$ is the action space, $\mathcal{P}$ represents the state transition probabilities, $R$ is the reward function, and $\gamma$ is the discount factor~\cite{Paul2025}. In the following, we describe these key components for the considered multi-tier satellite communication system.

\begin{enumerate}
    \item \textbf{State Space:} The state at any time slot $t$ captures the important system variables that influence decision-making but are not directly controlled through actions. These variables provide essential information about the current system condition to facilitate optimal decision-making. In the considered multi-tier satellite communication system, the state $s[t] \in \mathcal{S}$ is defined as follows:
    \begin{align}
    s[t]=\left\{t, s_t, N_t, \mathbf{D}(t), \mathbf{H}(t) \right\},
    \end{align}
    \noindent where $t$ is the current time step, $s_t$ is the index of the currently selected satellite, $N_t$ is the count of satellite handovers, $\mathbf{D}(t) = [D_1(t), D_2(t), \ldots, D_{N_C}(t)]$ represents the cumulative data delivered to each ground cluster, and $\mathbf{H}(t)$ captures the channel conditions between the HAP and all users in all clusters. These state variables provide the necessary context for making informed decisions.

    \item \textbf{Action Space:} The action space should directly correspond to the decision variables defined in our optimization problem (P1). As such, the action at time slot $t$, denoted as $a[t] \in \mathcal{A}$, consists of three components that align with the considered decision variables, \textit{i.e.},
    \begin{align}
    a[t]=(s_{t+1}, \mathbf{n}(t+1), \mathbf{u}(t+1)),
    \end{align}
    \noindent where $s_{t+1} \in \mathcal{L}_{t+1}$ represents the selected satellite for the next time slot (corresponding to decision variable $\boldsymbol{S}$), $\mathbf{n}(t+1) = [n_1(t+1), n_2(t+1), \ldots, n_{N_C}(t+1)]$ represents the subcarrier allocation to each ground cluster (corresponding to decision variable $\boldsymbol{N}$), and $\mathbf{u}(t+1) = [u_{1,t+1}, u_{2,t+1}, \ldots, u_{N_C,t+1}]$ represents the selected user from each cluster (corresponding to decision variable $\boldsymbol{U}$). 
    
    \item \textbf{Reward Function:} We design the reward function to directly reflect the optimization objectives of the formulated problem (P1). As such, the reward function should establish a clear correspondence between the immediate rewards received by the agent and our dual objectives of maximizing data throughput and minimizing satellite handover frequency. Accordingly, the reward function is defined as follows:
    \begin{align}
    r[t] = \eta \cdot R_{\text{total}}(t)  - \zeta \cdot \mathbb{I}(s_t \neq s_{t+1}),
    \end{align}
    where $R_{\text{total}}(t)$ represents the effective total downlink transmission rate at time $t$ (aligning with the objective $f_1$), and $\mathbb{I}(s_t \neq s_{t+1})$ is an indicator function that equals 1 when a satellite handover occurs (aligning with the objective $f_2$). Moreover, $\eta$ and $\zeta$ are weighting factors that control the trade-off between the two objectives. 
\end{enumerate}

\par By formulating the problem as an MDP, the DRL-based optimization framework can effectively maximize the cumulative reward through state-dependent action selection. In the following section, we present the specific algorithm that enables our framework to learn policies across the considered dynamic communication environment.

%
%
\subsection{Standard Truncated Quantile Critics (TQC) Algorithm}

\par To address the challenges posed by highly variable rewards in our multi-tier satellite communication optimization problem, we introduce the TQC algorithm~\cite{Kuznetsov2020}, which represents a significant advancement over standard soft actor-critic (SAC)~\cite{Haarnoja2018}. Specifically, TQC extends distributional reinforcement learning principles to effectively handle environments with high variance in returns and complex reward structures. By explicitly modeling the distribution of Q-values rather than just their expectations, TQC provides superior performance in scenarios where reward uncertainty is substantial~\cite{Xiao2023}, making it particularly suitable for the considered dynamic satellite communication system.

\par In particular, the innovation of TQC lies in its approach to representing and updating value functions. Different from the standard SAC, which maintains point estimates of Q-values, TQC models the entire distribution of returns using a set of quantile critics. This distributional perspective allows for more nuanced value estimation and uncertainty quantification~\cite{Kuznetsov2020}. Moreover, TQC prevents Q-value overestimation by systematically truncating the highest quantiles during the Bellman backup operation. This truncation mechanism effectively reduces optimistic bias in value estimates, leading to more stable and reliable policy learning.

\par Mathematically, TQC represents the distribution of the action-value function using $N$ quantiles. For a given state-action pair $(s[t], a[t])$, the quantile critic networks output $\{Z_i(s[t], a[t])\}_{i=1}^N$, representing the quantiles of the return distribution. As such, the policy update objective in TQC can be expressed as
\begin{equation}
\begin{aligned}
    J_{\text{policy}}(\theta) = \mathbb{E}_{s[t] \sim \mathcal{D}} \Big[ \mathbb{E}_{a[t] \sim \pi_\theta} [ \alpha \log \pi_\theta(a[t] | s[t]) \\ - \frac{1}{N-k}\sum_{i=1}^{N-k} Z_i^{\text{sorted}}(s[t], a[t]) ] \Big],
\end{aligned}
\label{eq:tqc_policy_objective}
\end{equation}

\noindent where $Z_i^{\text{sorted}}$ represents the sorted quantile estimates, and $k$ is the truncation parameter that determines how many of the highest quantiles to discard. This truncation is a key element that differentiates TQC from other distributional DRL algorithms.

\par Following this, the critic networks in TQC are updated using the quantile regression loss, which is defined as follows:
\begin{equation}
\begin{aligned}
L_Z(\phi) & = 
\\ & \mathbb{E}_{(s[t],a[t],r[t],s[t+1]) \sim \mathcal{D}} \Big[ \sum_{i=1}^{N-k} \rho_{\tau_i}(y[t] - Z_i(s[t], a[t])) \Big],
\end{aligned}
\label{eq:tqc_critic_loss}
\end{equation}
\noindent where $\rho_{\tau_i}$ is the quantile Huber loss function, $\tau_i$ represents the quantile fractions, and $y[t]$ is the target, \textit{i.e.},
\begin{equation}
\begin{aligned}
y[t]  = r[t] + &\gamma \Big( \frac{1}{N-k}\sum_{i=1}^{N-k} Z_i^{\text{sorted}}(s[t+1], a[t+1]) \\ 
&- \alpha \log \pi_\theta(a[t+1] | s[t+1]) \Big).
\end{aligned}
\label{eq:tqc_target}
\end{equation}
\noindent Note that the temperature parameter $\alpha$ in TQC serves the same purpose as in SAC, balancing exploration and exploitation. However, the entropy regularization is applied to a more robust estimate of action values, making the overall policy learning more stable in environments with highly variable rewards.

\par While TQC offers significant advantages for the considered satellite communication optimization problem, several critical limitations need to be addressed when applying it in solving our problem. \textbf{\textit{Firstly}}, the designed action space contains a substantial number of invalid satellite selection options (satellites that are unavailable or ineffective in certain states), which will lead to numerous inefficient action combinations. However, standard TQC lacks mechanisms to handle such constraints, thus causing the algorithm to waste computational resources exploring useless actions and significantly hampering convergence. \textbf{\textit{Secondly}}, the reward function confronts channel-induced volatility, which brings excessive noise to the assessment of policy value, thus making it difficult for TQC to achieve stable convergence with static hyperparameters. Therefore, we aim to propose specific improvements for TQC to overcome these limitations.

%
\subsection{LLM-guided TQC Algorithm with Dynamic Action Masking}

\par To address the limitations of standard TQC for the formulated problem, we propose LLM-guided TQC with dynamic action masking (LTQC-DAM), a novel algorithm that incorporates two key enhancements. \textit{Firstly}, we aim to propose a dynamic action masking mechanism that effectively constrains the action space to valid satellite options, thus eliminating unnecessary exploration of impossible actions while maintaining sufficient exploration through an adaptive noise strategy. \textit{Secondly}, we leverage LLMs to dynamically optimize TQC hyperparameters throughout the training process, addressing the channel-induced reward volatility.

\par
\begin{figure*}
  \centering
  \includegraphics[width=6.55 in]{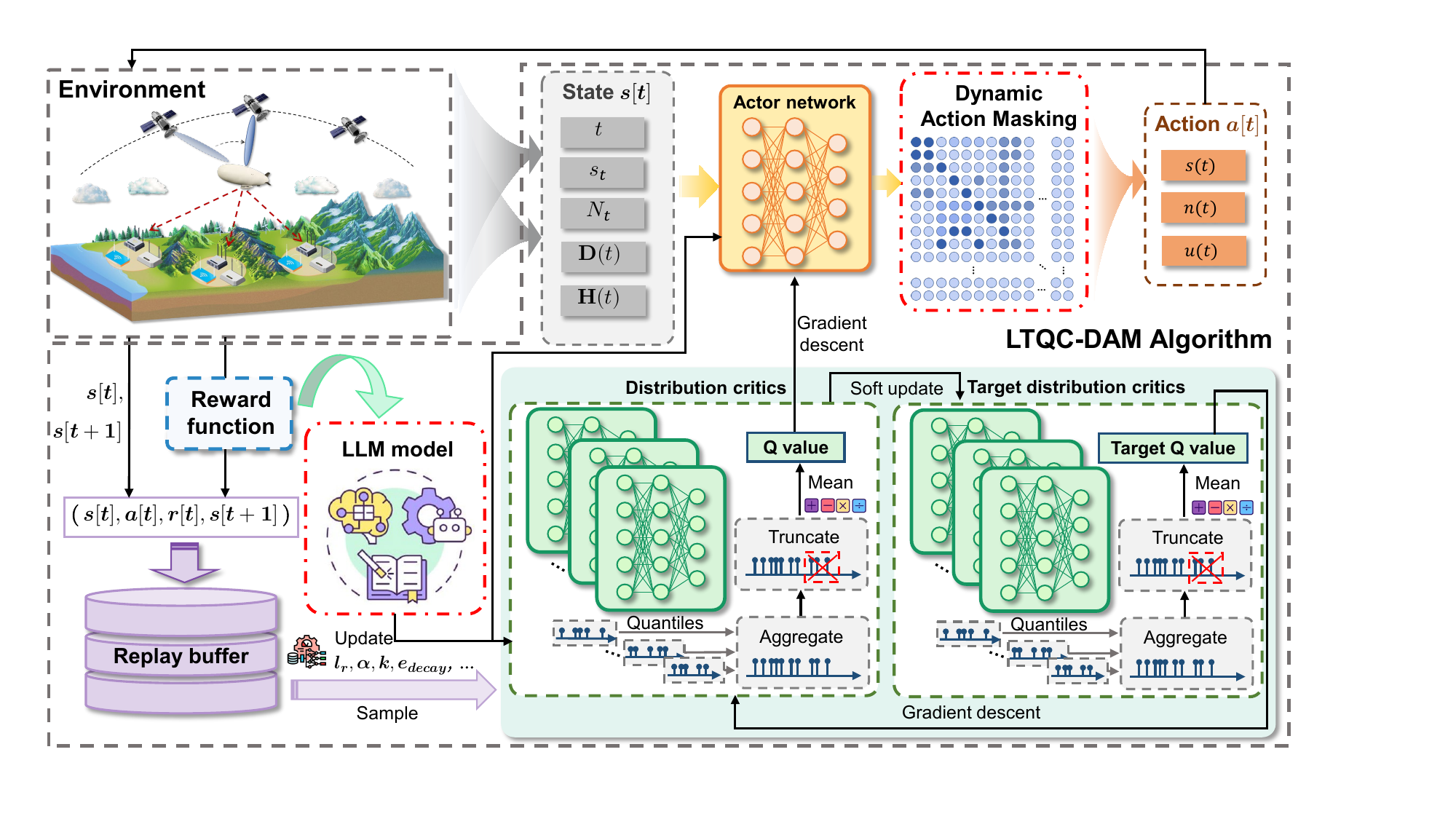}
  \caption{Framework of the LTQC-DAM algorithm for the considered multi-tier hybrid satellite downlink communication system.}
  \label{fig:algorithm}
\end{figure*}

 Fig.~\ref{fig:algorithm} illustrates the architecture of the proposed LTQC-DAM algorithm, which highlights the interaction between the dynamic action masking mechanism, the core TQC learning components, and the LLM-guided hyperparameter controller. The details of the dynamic action masking and LLM-guided adaptive hyperparameter tuning mechanisms are as follows. 

%

%
%
\subsubsection{Dynamic Action Masking with Adaptive Exploration}

\par The MDP of the considered multi-tier satellite downlink optimization problem presents a challenging action space for standard TQC, where only a subset of satellites is visible at any given timestep $t$. This constraint creates a high proportion of invalid actions that standard TQC cannot intrinsically recognize, which may lead to wasted computational resources and slower convergence as the algorithm attempts to learn from meaningless state-action pairs~\cite{Huang2022}. To address this limitation, we propose a dynamic action masking mechanism with adaptive exploration noise that guides the learning process toward valid satellite selections while maintaining sufficient exploration capability.

\par Specifically, for each timestep $t$, we define a binary visibility mask $M_t \in \{0,1\}^{|\mathcal{S}|}$ over the set of all satellites $\mathcal{S}$~\cite{Li2024a}, \textit{i.e.},
\begin{equation}
M_t[i] = 
\begin{cases}
1 & \text{if satellite } i \text{ is visible at time } t, \\
0 & \text{otherwise}.
\end{cases}
\label{eq:visibility_mask}
\end{equation}
\noindent Then, this mask is applied during the action selection process of the policy to restrict the output distribution to only valid satellites, \textit{i.e.,}
\begin{equation}
\pi_\theta(a|s[t], M_t) = \frac{\pi_\theta(a|s[t]) \cdot M_t[a]}{\sum_{a' \in \mathcal{S}} \pi_\theta(a'|s[t]) \cdot M_t[a']}.
\label{eq:masked_policy}
\end{equation}

\par Moreover, to ensure sufficient exploration despite the constrained action space, we implement an episode-dependent adaptive exploration mechanism. In particular, the probability of selecting an alternative satellite is governed by
\begin{equation}
\epsilon(e) = \max\left(\epsilon_0 \cdot \left(1 - \frac{e}{e_{decay} \cdot E}\right), 0\right),
\label{eq:epsilon_decay}
\end{equation}
\noindent where $\epsilon_0$ is the initial noise rate, $e$ is the current episode, $E$ is the total number of episodes, and $e_{decay}$ is a hyperparameter controlling the decay schedule. When triggered with probability $\epsilon(e)$, the mechanism randomly selects from the set of visible satellites excluding the current selection, which is given by
\begin{equation}
a[t] = 
\begin{cases}
a_{p}, \quad \quad \quad \quad \quad \quad \quad \quad \quad \text{with probability } 1-\epsilon(e), \\
\text{random}(\{i \in \mathcal{S} \mid M_t[i] = 1 \land i \neq a_{p}\}), \quad \text{otherwise},
\end{cases}
\label{eq:action_selection}
\end{equation}
\noindent where $a_{p}$ is the action initially selected by the policy.

\par By eliminating the evaluation of invalid actions, this mechanism helps the algorithm focus computational resources exclusively on meaningful learning opportunities within the feasible action space. Moreover, as training progresses, the adaptive exploration noise facilitates discovery of optimal handover strategies by promoting exploration of diverse yet valid satellite options, thus allowing the agent to identify effective transition patterns between satellites. 

%
%
\subsubsection{LLM-guided Adaptive Hyperparameter Optimization}

\par While TQC is effective for complex control tasks~\cite{Kuznetsov2020}, the algorithm suffers from sensitivity to hyperparameter configurations that often require extensive manual tuning. Such a challenge is particularly important in handling the MDP of the considered multi-tier satellite downlink optimization problem. This is due to the fact that the non-stationary environment, complex state-action dependencies, and sparse rewards create a landscape where optimal hyperparameters may vary throughout the training process~\cite{Bing2023}. In this case, static hyperparameters can lead to premature convergence or unstable learning dynamics, especially when the agent transitions between exploration and exploitation phases.

\par To address this limitation, we propose an LLM-guided adaptive hyperparameter tuning mechanism that leverages a large language model as a meta-controller for dynamic hyperparameter adjustment~\cite{Cheng2025}. This framework treats hyperparameter optimization as a sequential decision-making process where parameters are updated based on recent learning performance.

\par Specifically, the LLM-guided adaptive hyperparameter tuning mechanism can be formally described as follows. Let $\Theta_e = \{\theta_1^e, \theta_2^e, \ldots, \theta_{N_\theta}^e\}$ represent the set of hyperparameters at training iteration $e$. Then, we define a meta-optimization function $\mathcal{F}_{\text{LLM}}$ that maps the current training state to updated hyperparameters, \textit{i.e.},
\begin{equation}
\Theta_{e+\Delta e} = \mathcal{F}_{\text{LLM}}(\Theta_e, \mathcal{H}_e, \mathcal{P}_e),
\label{eq:llm_hyperparameter_update}
\end{equation}
\noindent where $\mathcal{H}_e = \{r[e-k], r[e-k+1], \ldots, r[e]\}$ represents a window of $k$ recent episode rewards, and $\mathcal{P}_e = \frac{e}{E}$ represents the normalized training progress, where $E$ is the total number of training episodes.

\par Following this, the LLM is prompted with structured information, which is given by
\begin{equation}
\begin{aligned}
\text{prompt} = 
\begin{bmatrix}
\Theta_e &= \{\theta_1^e, \theta_2^e, \ldots, \theta_{N_\theta}^e\} \\
\mathcal{H}_e &= \{r[e-k], r[e-k+1], \ldots, r[e]\} \\
\mathcal{P}_e &= \frac{e}{E}
\end{bmatrix},
\end{aligned}
\label{eq:llm_prompt}
\end{equation}
\noindent where $\Theta_e$ contains the hyperparameters such as discount factor, entropy coefficient, learning rate, batch size, soft update coefficient, the number of quantiles, and the decay schedule $e_{decay}$. For each parameter $\theta_i$, the LLM optimizes within constrained bounds, \textit{i.e.},
\begin{equation}
\theta_i^{e+\Delta e} \in [\theta_i^{\text{min}}, \theta_i^{\text{max}}].
\label{eq:parameter_bounds}
\end{equation}

\par As such, the LLM analyzes reward patterns to determine appropriate parameter adjustments~\cite{Huang2025}. For example, when observing oscillating rewards, it might reduce the learning rate to stabilize learning; when rewards plateau, it may increase exploration by adjusting the decay schedule $e_{decay}$.

\par Accordingly, this mechanism benefits from the rich domain expertise encoded within the knowledge base of LLM about DRL algorithm behavior, thus effectively transferring insights about parameter sensitivities and optimal configurations from a vast corpus of training data to the considered optimization problem. As such, we can detect and respond intelligently to various training pathologies such as catastrophic forgetting or reward plateaus, and then implement corrective parameter adjustments that help overcome these challenges. By incorporating the mechanism, the algorithm will adapt dynamically to the non-stationary nature of the learning process.

%
%
\subsubsection{The Main Steps and Complexity Analyses of the Proposed LTQC-DAM}

\par The proposed LTQC-DAM algorithm operates in a two-tier optimization process where the action masking component focuses on immediate efficiency by filtering invalid satellite selections at each decision step, while the LLM-guided hyperparameter optimization provides meta-level control that adapts algorithm behavior based on observed learning dynamics. As such, the main steps of the LTQC-DAM algorithm are shown in Algorithm~\ref{alg:ltqc-dam}, and the complexity analyses are detailed as follows. 

\par Specifically, we analyze the computational and space complexity of the proposed LTQC-DAM algorithm for both the training and execution phases.

\par \textbf{Training Phase:} The computational complexity of LTQC-DAM can be expressed as $\mathcal{O}(|\theta| + N|\phi| + ETG + ET|\theta| + ET(N|\phi|) + \frac{E}{\Delta e}C_{LLM})$, which can be broken down as follows:

\begin{itemize}
    \item \textit{\textbf{Network Initialization:}} The computational complexity for initializing network parameters is $\mathcal{O}(|\theta| + N|\phi|)$~\cite{Kuznetsov2020}, where $|\theta|$ represents the number of parameters in the policy network, and $|\phi|$ denotes the average number of parameters in each of the $N$ quantile critic networks.

    \item \textit{\textbf{Replay Buffer Collection:}} The complexity of collecting state transitions in the replay buffer is $\mathcal{O}(ETG)$~\cite{Zhang2025}, where $E$ is the total number of training episodes, $T$ represents the number of steps in each episode, and $G$ is the complexity of environment interactions, including the generation of visibility masks as described in Eq.~\eqref{eq:visibility_mask}.

    \item \textit{\textbf{Network Updates:}} This phase consists of policy network updates and quantile critic network updates~\cite{Zhang2025}. The policy network update based on Eq.~\eqref{eq:tqc_policy_objective} has complexity $\mathcal{O}(ET|\theta|)$, while updating the $N$ critic networks based on Eq.~\eqref{eq:tqc_critic_loss} has complexity $\mathcal{O}(ET(N|\phi|))$.
  
    \item \textit{\textbf{LLM-guided Hyperparameter Tuning:}} The hyperparameter tuning occurs every $\Delta e$ episodes and has complexity $\mathcal{O}(\frac{E}{\Delta e}C_{LLM})$, where $C_{LLM}$ is the computational cost of a single LLM inference operation for hyperparameter optimization based on Eq.~\eqref{eq:llm_hyperparameter_update}.
\end{itemize}

\par The space complexity of LTQC-DAM accounts for the neural network parameters, replay buffer, and LLM-related storage. This can be expressed as $\mathcal{O}(|\theta| + N|\phi| + |\mathcal{D}|(2|s| + |a| + 1 + |M|) + |H_e|)$, where $|\mathcal{D}|$ represents the replay buffer size, $|s|$ and $|a|$ denote the dimensions of state and action spaces respectively, $|M|$ is the size of the visibility mask, and $|H_e|$ is the size of the reward history window used for LLM-guided optimization.

\par \textbf{Execution Phase:} During execution, the computational complexity of LTQC-DAM is $\mathcal{O}(T(|\theta| + |M|))$, which includes the generation of visibility masks according to Eq.~\eqref{eq:visibility_mask} and action selection from the masked policy distribution as per Eq.~\eqref{eq:masked_policy}~\cite{Zhang2025}. The space complexity is $\mathcal{O}(|\theta| + |M|)$, primarily from storing the policy network parameters and the visibility mask.

\par Notably, the dynamic action masking mechanism adds a minor computational overhead of $\mathcal{O}(|M|)$ per step but significantly improves sample efficiency by restricting the exploration to valid actions only. The LLM-guided hyperparameter tuning introduces a periodic computational cost of $\mathcal{O}(C_{LLM})$ but considerably enhances training stability and convergence speed. Thus, it makes a worthwhile trade-off for the formulated optimization problem with high reward variance.

\begin{algorithm}[t]
\small
    \caption{LLM-guided TQC with Dynamic Action Masking (LTQC-DAM)}
    \label{alg:ltqc-dam}
    \textbf{Initialize:} Replay buffer $\mathcal{D}$, policy network parameters $\theta$, critic network parameters $\phi$, $N$ quantiles, truncation parameter $k$, temperature parameter $\alpha$, initial hyperparameters $\Theta_0$\;
    \For{episode $e=1,2,\cdots, E$}{
        Initialize environment, obtain initial state $s[0]$\;
        \For{$t=0,1,\cdots, T-1$}{
            Generate visibility mask $M_t$ according to Eq.~\eqref{eq:visibility_mask}\;
            Sample action from masked policy distribution: $a_{policy} \sim \pi_\theta(a|s[t], M_t)$ using Eq.~\eqref{eq:masked_policy}\;
            Apply adaptive exploration with probability $\epsilon(e)$ from Eq.~\eqref{eq:epsilon_decay} to obtain final action $a[t]$ according to Eq.~\eqref{eq:action_selection}\;
            Execute action $a[t]$, observe reward $r[t]$ and next state $s[t+1]$\;
            Store transition $(s[t], a[t], r[t], s[t+1], M_t)$ in replay buffer $\mathcal{D}$\;
          
            \If{update step}{
                Sample mini-batch of transitions from $\mathcal{D}$\;
                \For{each critic network}{
                    Compute target $y[t]$ using Eq.~\eqref{eq:tqc_target}\;
                    Update critic parameters $\phi$ by minimizing the loss in Eq.~\eqref{eq:tqc_critic_loss}\;
                }
                Update policy parameters $\theta$ by minimizing the objective in Eq.~\eqref{eq:tqc_policy_objective}\;
                Update target network parameters using soft update\;
            }
        }
      
        Store episode reward $r[e] = \sum_{t=0}^{T-1} r[t]$\;
      
        \If{$e$ mod $\Delta e = 0$}{
            Collect reward history $\mathcal{H}_e = \{r[e-k], r[e-k+1], \ldots, r[e]\}$\;
            Calculate normalized training progress $\mathcal{P}_e = \frac{e}{E}$\;
            Construct LLM prompt using Eq.~\eqref{eq:llm_prompt}\;
            Update hyperparameters via LLM: $\Theta_{e+\Delta e} = \mathcal{F}_{\text{LLM}}(\Theta_e, \mathcal{H}_e, \mathcal{P}_e)$ using Eq.~\eqref{eq:llm_hyperparameter_update}\;
            Ensure that all parameters remain within bounds according to Eq.~\eqref{eq:parameter_bounds}\;
            Apply updated hyperparameters (learning rate, $\alpha$, $k$, $e_{decay}$, etc.)\;
        }
    }
\end{algorithm}

%
%
\section{Simulation Results and Analysis}
\label{sec:simulation_results_and_analysis}

\par In this section, we conduct key simulations to evaluate the performance of the proposed LTQC-DAM method for solving the formulated optimization problem.

\subsection{Simulation Setups}

%
%
\subsubsection{Scenario and Algorithm Setups}


\begin{figure*}
    \centering
    \subfloat[Rewards of different baselines]{
    \includegraphics[width=0.33\linewidth]{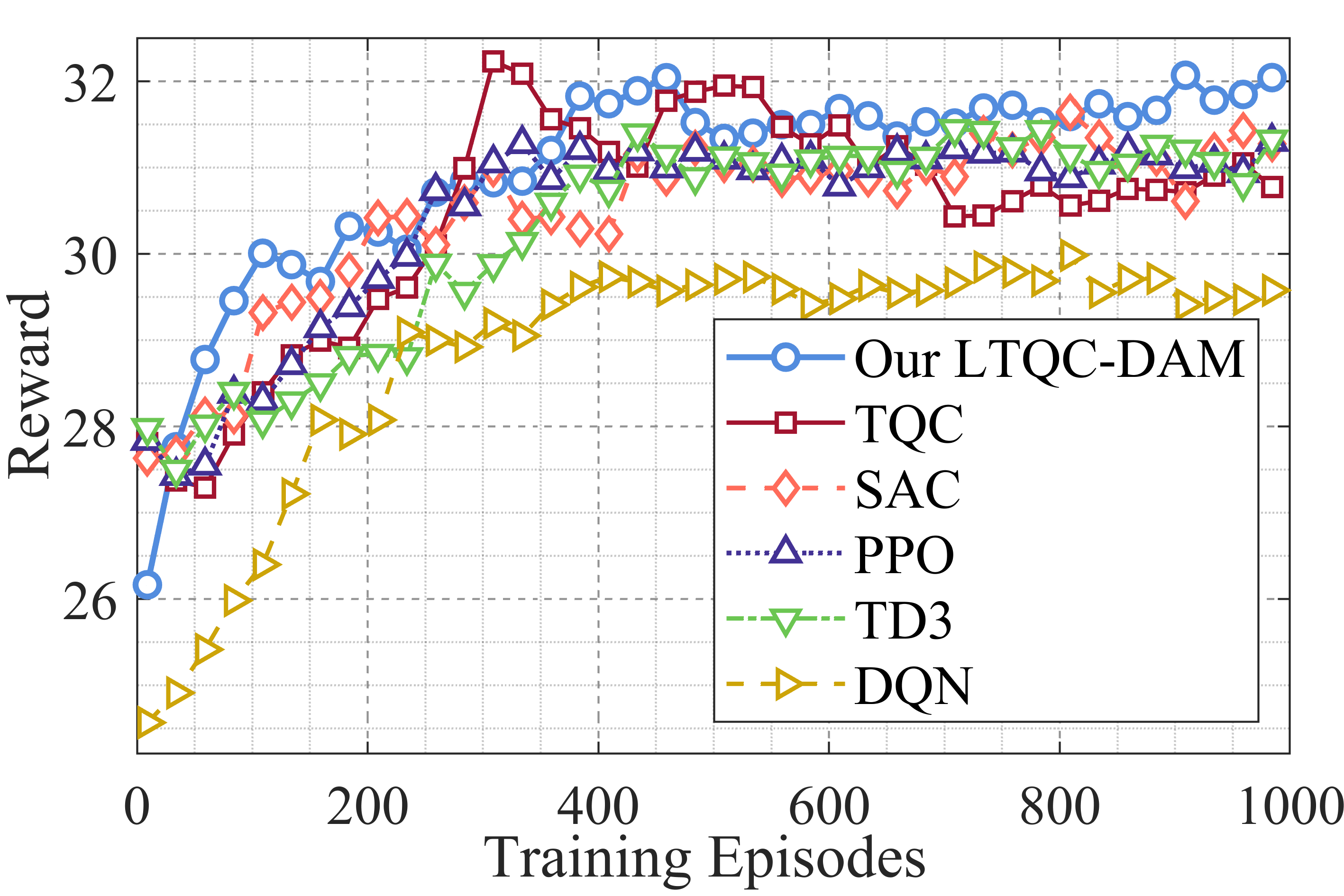}\label{fig:reward1}}
    \subfloat[Handover frequency of different baselines]{
    \includegraphics[width=0.33\linewidth]{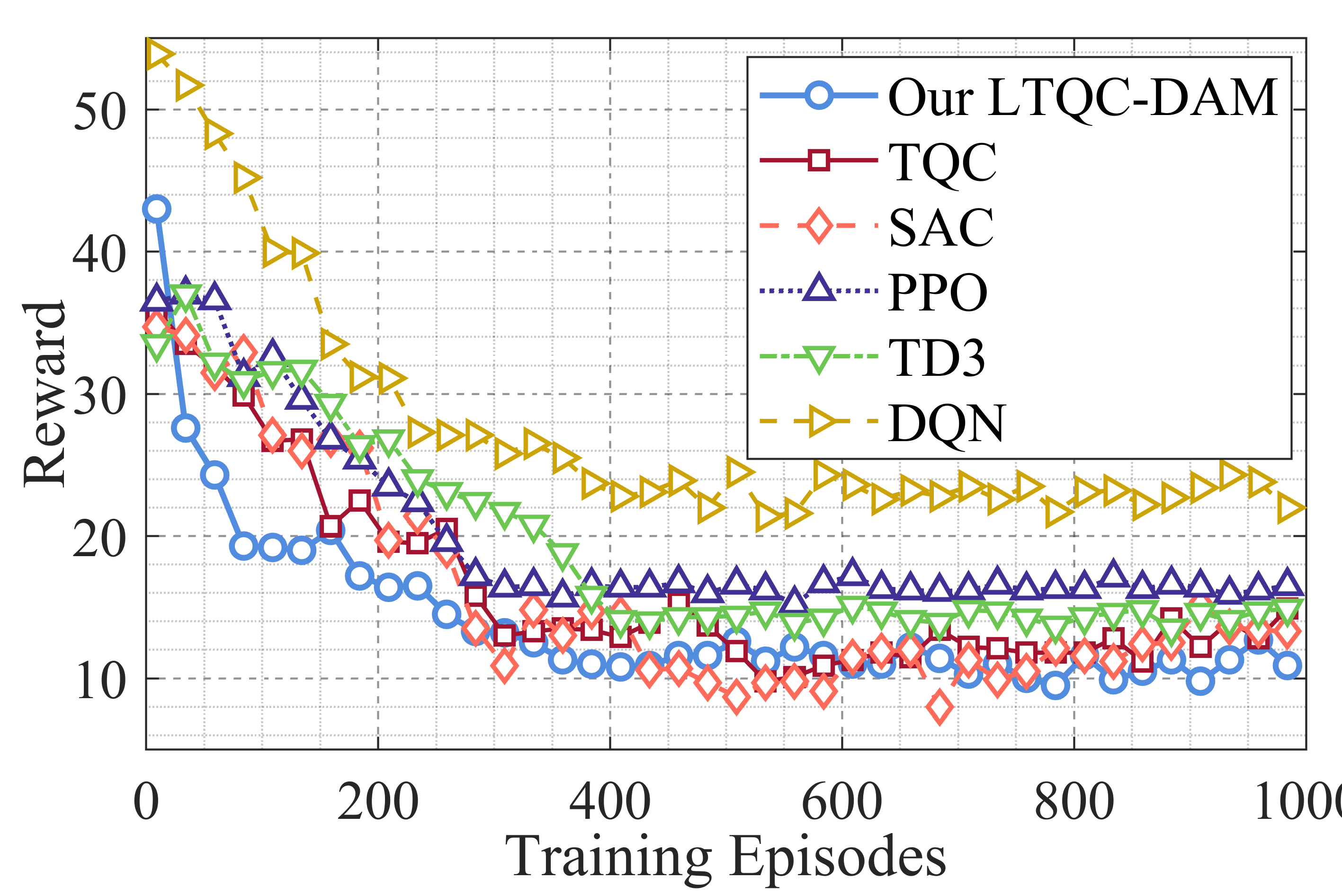}\label{fig:switch}} 
    \subfloat[Rewards of different LLMs]{
    \includegraphics[width=0.33\linewidth]{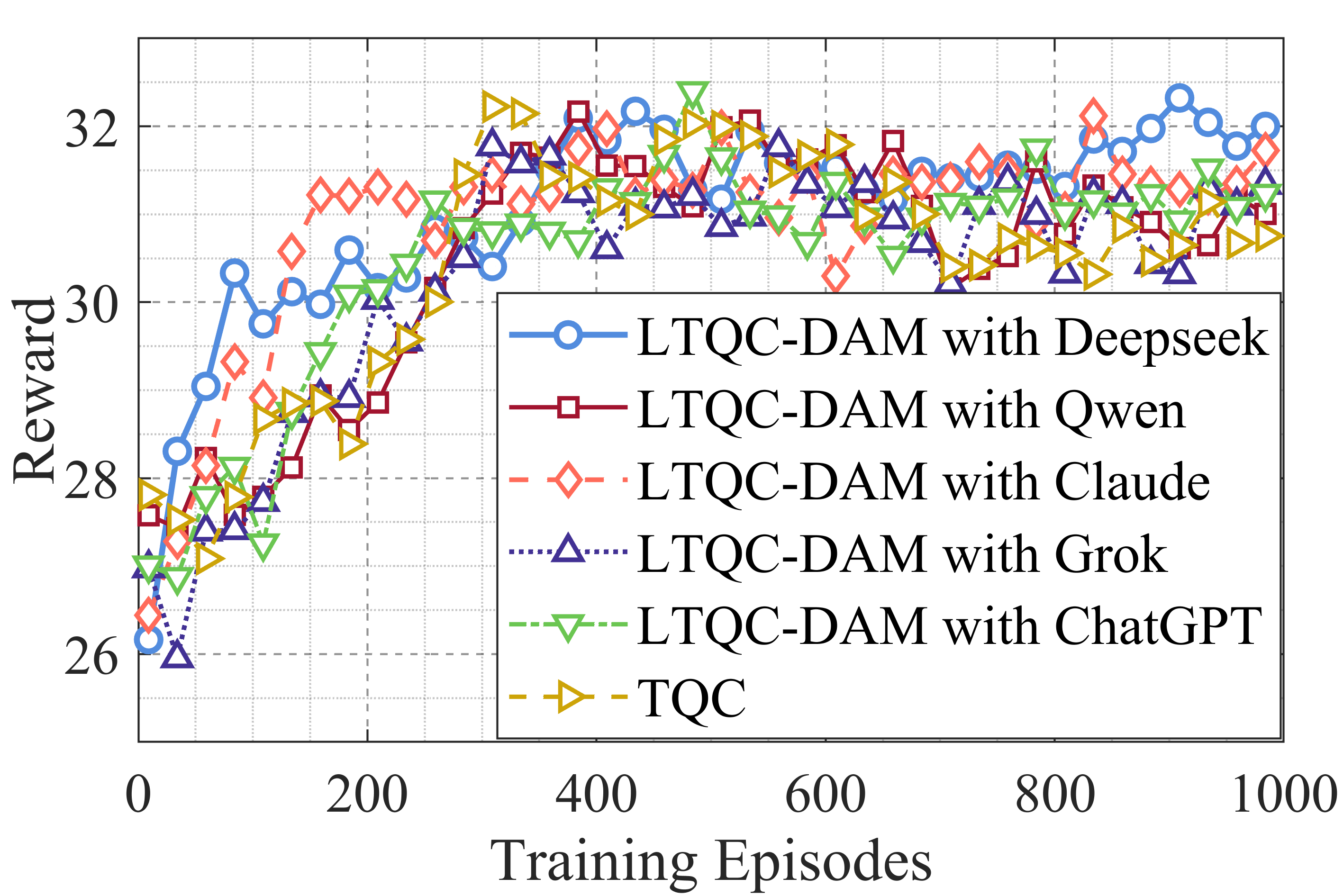}\label{fig:reward2}} 
       \\
  \caption{Convergence comparison of different algorithms.} 
  \label{fig: convergence}
\end{figure*}

\par We consider a multi-tier hybrid satellite downlink communication system consisting of a LEO satellite constellation, a HAP, and multiple ground user clusters. Specifically, the constellation comprises 110 LEO satellites, of which 80 are positioned at an altitude of $5 \times 10^5$ m and 30 at $10^6$ m. Most satellites orbit near the equatorial plane, while some maintain inclination angles of approximately $\pm \pi / 8$. Satellites sharing the same orbital plane are evenly distributed throughout their respective orbits~\cite{Okati2020,Deng2021}. Moreover, the HAP operates at an altitude of 20000 m and follows the Gauss-Markov mobility model shown in Section~\ref{ssec: HAP}. In addition, we consider three geographically dispersed user clusters, each capable of intra-cluster communication, with randomly distributed users within each cluster. Following this, the communication link between satellites and the HAP utilizes FSO, while RF links connect the HAP to the ground user clusters, and these communication-related parameters follow~\cite{Wu2024}. 

\par The proposed LTQC-DAM algorithm uses a policy network with three fully connected layers (256, 256, 128 neurons) and ReLU activation functions. Moreover, the discount factor, learning rates, truncation parameter, soft update coefficient, and decay schedule parameters are set as 0.999, 1e-4, 2, 0.005, and 0.3 initially. 

\par Note that these hyperparameters may be adjusted dynamically during training through LLM guidance. Note that we introduce multiple LLMs accessible through public APIs. In particular, the primary model used is DeepSeek V3, which serves as the default optimization engine. Additionally, we incorporate ChatGPT 4.1, Claude 3.7 Sonnet, Grok 3, and Qwen 2.5 to provide diverse optimization perspectives. Each LLM may offer unique insights based on its particular strengths in understanding reinforcement learning dynamics and optimization strategies. Each training episode consists of 60 steps, and the algorithm was trained for a total of 1000 episodes to ensure convergence.

%
\subsubsection{Baseline Algorithms}

\par To evaluate the performance of the proposed LTQC-DAM algorithm, we introduce several DRL methods as baselines, including the following algorithms. 

\begin{itemize}
    \item \textit{SAC}~\cite{Haarnoja2018}: An off-policy actor-critic algorithm that maximizes both expected return and entropy, offering excellent sample efficiency and stability.
  
    \item \textit{Twin Delayed Deep Deterministic Policy Gradient (TD3)}~\cite{Dankwa2019}: A deterministic policy gradient algorithm that addresses function approximation errors by employing twin critics and delayed policy updates.
  
    \item \textit{Proximal Policy Optimization (PPO)}~\cite{Schulman2017}: An on-policy algorithm that optimizes policies via a surrogate objective function while constraining policy changes to improve stability.
  
    \item \textit{Deep Q-Network (DQN)}~\cite{Fan2020}: A value-based algorithm using deep neural networks to approximate Q-values, with experience replay and target networks for stability.
  
    \item \textit{Standard TQC}~\cite{Kuznetsov2020}: The standard TQC algorithm without LLM-guided hyperparameter tuning, which uses distributional critics to model return distributions and truncates the highest quantiles to prevent overestimation.
\end{itemize}

\par Note that all baseline algorithms are implemented with the same dynamic action masking mechanism as the proposed method for fair comparison, thereby ensuring that they only explore valid actions in the satellite communication environment.

%
%
\subsection{Performance Evaluation}

\subsubsection{Convergence Analyses}

\par Fig.~\ref{fig: convergence}(a) illustrates the convergence behavior of the proposed LTQC-DAM algorithm compared to baseline algorithms in terms of episodic reward. As can be seen, LTQC-DAM demonstrates significantly faster convergence and achieves higher final rewards. Moreover, the standard TQC algorithm converges more rapidly than other algorithms but exhibits significant fluctuations in later stages. This may be primarily due to its sensitivity to hyperparameters, where static parameter settings prove inadequate across different training phases. In addition, DQN shows the weakest convergence performance, possibly because this value-based algorithm struggles to adapt to the reward volatility present in the considered problem. 

\par \par Fig.~\ref{fig: convergence}(b) illustrate the convergence behaviors for optimization objective $f_2$, \textit{i.e.},  minimizing satellite handover frequency. As can be seen, all the algorithms exhibit convergence for $f_2$. Moreover, the proposed LTQC-DAM achieves relatively rapid convergence and demonstrates stability post-convergence. Overall, all baseline algorithms achieve reasonable convergence performance due to the incorporation of the dynamic action masking mechanism. Moreover, the proposed LTQC-DAM algorithm exhibits superior performance, which can be attributed to the truncated quantile critic structure of TQC and the enhanced adaptability provided by LLM-based hyperparameter adjustment. Note that due to the frequently changing channel conditions, the optimization objective $f_1$ can exhibit oscillatory. Thus, we evaluate its average performance in the following. 

\subsubsection{Comparisons and Analyses}

\par
\begin{figure}
    \centering
    \includegraphics[width=3.5 in]{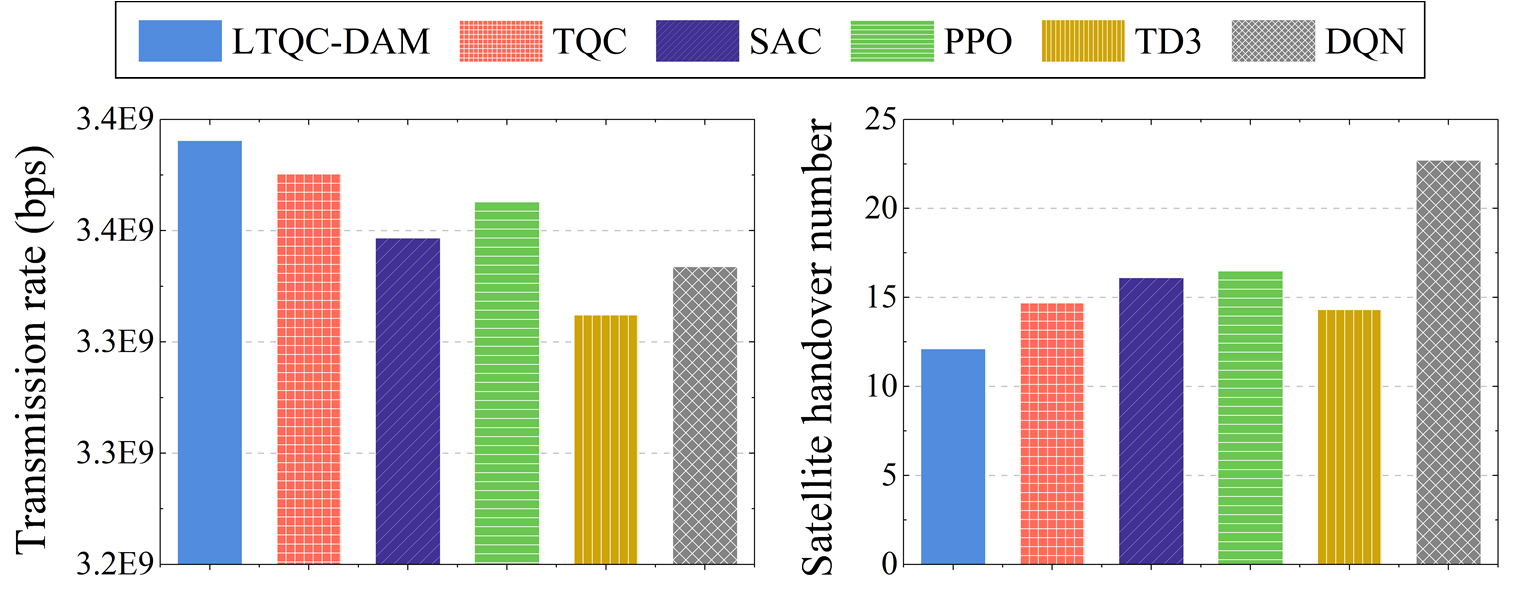}
    \caption{Average optimization objective performance comparison of different algorithms, including downlink transmission rate ($f_1$) and satellite handover frequency ($f_2$).}
    \label{fig:performance_comparison}
\end{figure}

 Fig.~\ref{fig:performance_comparison} illustrates the average objective values of the proposed algorithm and benchmark algorithms after convergence over 60 time slots. As can be seen, the proposed LTQC-DAM algorithm outperforms all benchmark algorithms across both metrics. Moreover, the proposed LTQC-DAM algorithm achieves a 0.44\% improvement in $f_1$ but a 17.69\% improvement in $f_2$ compared to the standard TQC algorithm. This superior performance can be attributed to the LLM-guided hyperparameter optimization, which ensures optimal adaptation to the specific characteristics of the multi-tier satellite downlink communication environment throughout the training process.

%
%
\subsection{Impact of Different LLM Models}

\begin{figure}
    \centering
    \includegraphics[width=3.5 in]{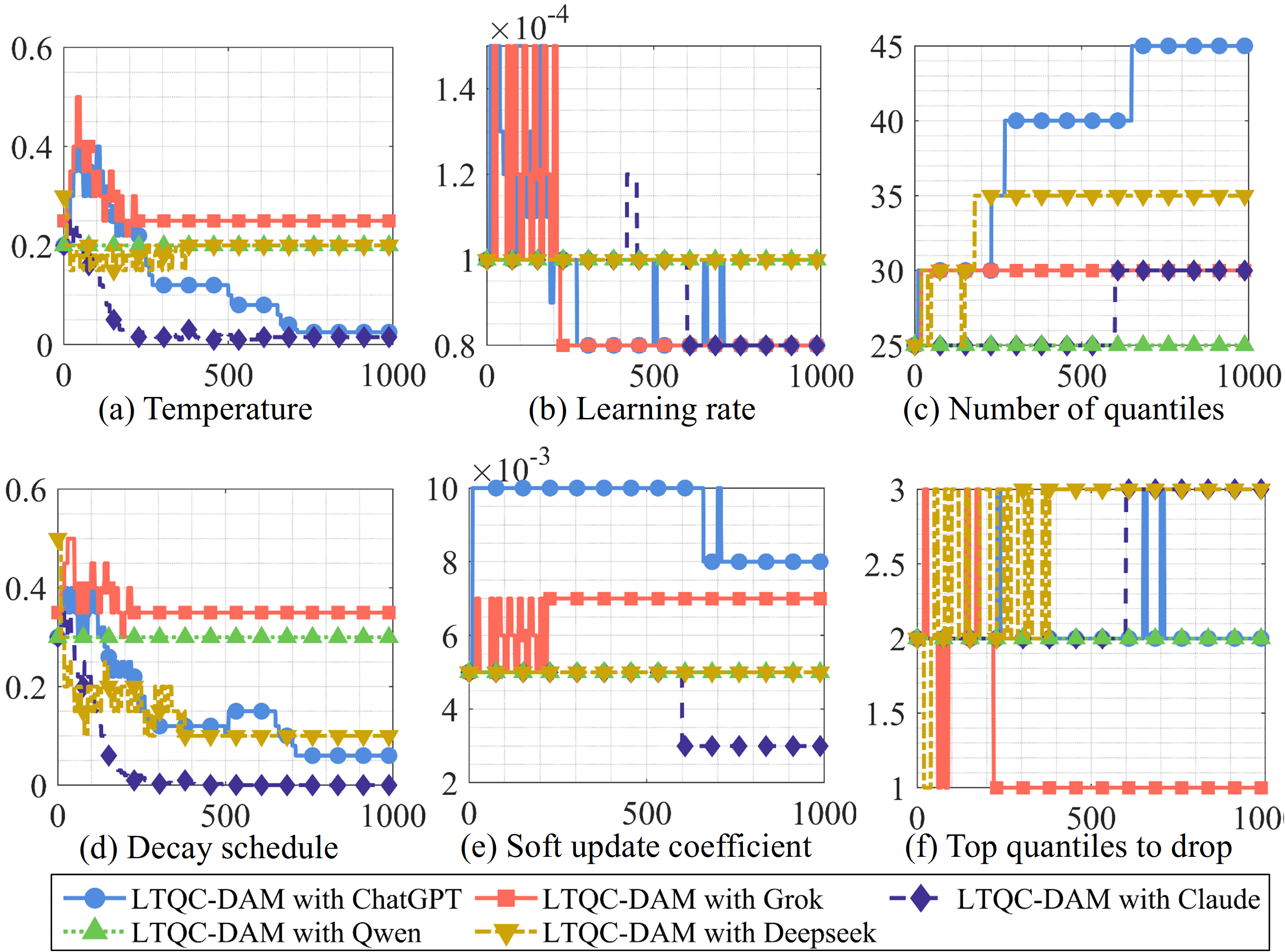}
    \caption{Visualization of hyperparameter adaptation patterns across different LLMs during training.}
    \label{fig:llm_comparison}
\end{figure}

\par To investigate the impact of different LLM models, we implement the proposed LTQC-DAM with five state-of-the-art LLMs, including DeepSeek, Qwen, Claude, ChatGPT, and Grok. Fig.~\ref{fig: convergence}(c) shows the convergence behaviors achieved by each variant after 1000 training episodes. As shown in the figure, DeepSeek consistently outperforms other LLMs by achieving higher and more stable rewards. Moreover, Claude and Qwen deliver comparable performance, ranking second and third, respectively, while ChatGPT and Grok show moderate but still beneficial improvements over the baseline TQC algorithm without LLM guidance.

\par To understand why DeepSeek produces superior results, we further visualize the hyperparameter adaptation patterns of different LLMs during training in Fig.~\ref{fig:llm_comparison}. From this visualization, we can derive three key observations and insights as follows.

\begin{enumerate}
    \item DeepSeek demonstrates more gradual and nuanced hyperparameter adjustments than the other LLM models, which tend to make larger, sometimes oscillatory changes. This methodical approach potentially avoids destabilizing the learning process while still providing sufficient adaptation, thereby facilitating the convergence.
  
    \item When analyzing the learning rate adaptation in Fig.~\ref{fig:llm_comparison}, DeepSeek keeps the learning rate unchanged during training stages, thereby ensuring a consistent learning environment. In contrast, other models such as ChatGPT and Grok show frequent and sometimes contradictory adjustments to critical parameters during later training stages, potentially causing policy instability.
  
    \item DeepSeek exhibits more context-awareness in the recommendations for parameters, often making correlated adjustments across multiple hyperparameters simultaneously. For instance, when increasing the temperature parameter for more exploration, it appropriately adjusts the truncation parameter to compensate for the increased variance, thus maintaining stable learning.
\end{enumerate}

\par Consequently, the analysis of hyperparameter trajectories reveals that the most effective LLM-guided optimization occurs when parameter adjustments are relatively small and guided by clear reasoning about the current learning stage of the algorithm. In this case, DeepSeek appears better aligned with such requirements. The reason may be attributed to its no auxiliary loss load balancing strategy, which dynamically adjusts expert loads through bias terms rather than disruptive loss functions. Meanwhile, the multi-head latent attention architecture of DeepSeek also enables more context-aware parameter recommendations by efficiently handling complex dependencies across training contexts.

%
%
\section{Conclusion} 
\label{sec:conclusion}

\par This paper has investigated a multi-tier hybrid FSO/RF satellite downlink communication system. Following this, we have formulated a joint optimization problem to simultaneously balance downlink transmission rate and handover frequency by optimizing network configuration and satellite handover decisions. The problem has proved highly challenging due to its dynamic and non-convex nature with time-coupled constraints and complex cross-tier dependencies between multi-tiers. To address these challenges, we have proposed a novel LTQC-DAM that incorporates dynamic action masking to eliminate unnecessary exploration and employs LLMs to adaptively tune hyperparameters. Simulation results have demonstrated that the proposed LTQC-DAM algorithm achieves faster convergence and superior performance compared to various baseline algorithms in terms of both downlink transmission rate and handover frequency. Furthermore, our comparative analysis of different LLM models has revealed that DeepSeek delivers the best performance through gradual and contextually aware parameter adjustments. Future work will focus on fine-tuning and customizing LLMs specifically for such systems, which may potentially yield even better performance by incorporating domain-specific knowledge.


\bibliographystyle{IEEEtran}
\bibliography{myref}

\end{document}